\documentclass[aps,prb,twocolumn,showpacs,tightenlines]{revtex4}
\usepackage{epsfig}
\tighten
\def \cT {{\cal T}}

\def \cG {{\cal G}}

\def \cT {{\cal T}}

\begin{document}
\title{Delocalization and conductance quantization in one-dimensional systems attached to leads}
\date{\today}
\author{Z. Y. Zeng$^{1,2}$ and F. Claro$^1$}
\affiliation{$^1$ Facultad de F\'isica, Pontificia Universidad
Cat\'olica de Chile, Casilla 306, Santiago 22, Chile\\
$^2$ Department of Physics  National University of Singapore, 10
Kent Ridge Crescent,  119260 Singapore}
\date{\today}

\begin{abstract}
We investigate the delocalization and conductance quantization in
finite one-dimensional chains with only off-diagonal disorder
coupled to leads. It is shown that the appearance of delocalized
states at the middle of the band under correlated disorder is
strongly dependent upon the even-odd parity of the number of sites
in the system. In samples with inversion symmetry the conductance
equals $2e^{2}/h$ for odd samples, and is smaller for even parity.
This result suggests that this even-odd behavior found previously
in the presence of electron correlations may be unrelated to
charging effects in the sample.
\end{abstract}

\maketitle

Since the pioneering work of Anderson,\cite{Anderson} localization in
disordered systems has become a key issue in solid state physics. Mott and
Twose\cite{Mott} suggested that all the electronic eigenstates in less than
two-dimensional disordered systems are localized. Borland\cite{Borland} gave
a rather general proof of this statement. Economou and Cohen\cite{Economou}
have re-examined the localization problem in the 1D tight-binding model,
concluding that all states are localized if and only if the nearest-neighbor
coupling is considered. However, Theodorou and Cohen\cite{Theodorou} showed
that the state at the middle of the band is extended, regardless of the
randomness of the nearest-neighbor hopping matrix elements. Recently, it has
also been argued that the delocalization transition exists in1D systems with
correlated diagonal and/or non-diagonal disorder, i.e., that at some
particular energies the states are extended.\cite{Hilke} The delocalization
transition has now been investigated in 1D random quantum Ising chains\cite
{Shankar}, 1D random XY models\cite{Mckenzie}, weakly disordered quasi 1D
tight-binding hopping models\cite{Brouwer1} and dirty superconducting wires.
\cite{Brouwer2}.

On the other hand, Oguri\cite{Oguri} found an even-odd parity effect in the
conductance characteristics of a finite Hubbard chain coupled with continuum
states. He attributed such a parity effect to the many-body Kondo resonance
and the presence of the reservoirs of continuum states, while Brouwer et al.%
\cite{Brouwer1} explained it by level repulsion of the transmission
eigenmodes. Similar even-odd behavior in the conductance has been found in
first principles numerical calculations in monatomic molecular wires within
the local-density-functional approximation, and interpreted as arising from
charge neutrality and resonant tunneling due to the sharp tip structure.\cite
{Sim}

Here we address the question of whether these effects are many
body effects or not. Our approach is to check if they arise in the
absence of a Coulomb term in the hamiltonian.Within a tight
binding model in the absence of electron-electron interactions we
find that both delocalization in a correlated disordered sample
and the even-odd parity feature in symmetric strings are present,
irrespective of any charging effects.

For the sake of simplicity, we consider only non-diagonal disorder keeping
on-site energies the same at all sites. We derive an explicit delocalization
condition to be satisfied by the parameters associated with hopping between
sites and coupling to the reservoirs of continuum states. We find that it
depends on the parity of the string, i.e., on whether the number of sites is
even or odd. Applied to the molecular wire structure with inversion
symmetry, we observe that the conductance in the odd case always equals $%
2e^{2}/h$ while it is smaller than this quantity if the number of sites is
even.

The system considered here is a chain of $N$ sites labelled $1,2,\cdots ,N$
\ from left (L) to right (R), with its ends connected to reservoirs with
chemical potential $\epsilon _{F}$. The hamiltonian is
\begin{eqnarray}
H&=&\sum\limits_{i=1}^{N}\epsilon
_{0}a_{i}^{+}a_{i}+\sum\limits_{i=1}^{N-1}(t_{i}a_{i}^{+}a_{i+1}+H.c.)+
\sum%
\limits_{k,r=L,R}\epsilon _{kr}b_{kr}^{+}\nonumber \\ & &b_{kr}+
\sum%
\limits_{k}(V_{k}^{L}b_{kL}^{+}a_{1}+V_{k}^{R}b_{kR}^{+}a_{N}+H.c.),
\end{eqnarray}
where $a_{i}$$(b_{k})$ is the annihilation operator of electron at site $i$
(lead $r$), and the other terms have their usual meaning. We will
characterize delocalization by perfect transmission at some particular
energy, i.e., by a transmission coefficient that equals $1$. The
corresponding state is then extended.\cite{Theodorou} Considering the sites
to the right of site $1$ as part of the right reservoir, the Keldysh
formalism\cite{Datta} yields for the transmission probability the expression,
\begin{widetext}
\begin{eqnarray}
&&  \nonumber \\
\cT(\epsilon_F) &=&\frac{-2\Gamma ^{L}|t_{1}|^{2}ImG_{2R}^{r}}{(\epsilon
_{F}-\epsilon _{0}-|t_{1}|^{2}ReG_{2R}^{r})^{2}+(\Gamma
_{L}-2|t_{1}|^{2}ImG_{2R}^{r})^{2}/4},
\end{eqnarray}

where
\begin{eqnarray*}
\Gamma ^{L/R} &=&\sum\limits_{k}2\pi |V_{k}^{L/R}|^{2}\delta
(\epsilon_F
-\epsilon _{kL/R}), \\
G_{iR}^{r}(\epsilon _{F}) &=&[g_{i}^{r}(\epsilon _{F}))^{-1}-\mid t_{i}\mid
^{2}G_{i+1,R}^{r}(\epsilon _{F})]^{-1},\hspace{1cm}i=2,3,\cdots ,N-1 \\
G_{NR}^{r}(\epsilon_F ) &=&[g(\epsilon
_{F}))^{-1}+\frac{i}{2}\Gamma ^{R}]^{-1}
\\
g_{i}^{r}(\epsilon _{F}) &=&(\epsilon _{F}-\epsilon _{0}+i0^{+})^{-1}.%
\hspace{1cm}i=1,2,\cdots N
\end{eqnarray*}
\end{widetext}
and the usual notation for Green functions has been employed. The
transmission probability depends in general on the position of the Fermi
level at the reservoirs as well as on the disorder configuration of the
hopping parameters $(t_{1},t_{2},\cdots ,t_{N-1})$. Here we consider the
most interesting case, when the Fermi level $\epsilon _{F}$ is pinned at the
value of the independent site energy $\epsilon _{0}$, the middle of the band
or level group of the 1D system.\cite{Hilke,Brouwer1} The real part of all
retarded Green functions becomes zero and the first term in the denominator
of Eq. (2) vanishes. Perfect transmission through the 1D lattice is then
obtained if
\begin{equation}
\Gamma ^{L}=-2|t_{1}|^{2}ImG_{2R}^{r},
\end{equation}
where now

\begin{eqnarray}
ImG_{2R}^{r} &=&-\frac{\Gamma ^{R}}{2}|\frac{t_{3}t_{5}\cdots t_{N-2}}{%
t_{2}t_{4}\cdots t_{N-1}}|^{2};\hspace{0.2cm}N\hspace{0.1cm}odd \\
ImG_{2R}^{r} &=&-\frac{2}{\Gamma ^{R}}|\frac{t_{3}t_{5}\cdots t_{N-1}}{%
t_{2}t_{4}\cdots t_{N-2}}|^{2};\hspace{0.2cm}N\hspace{0.1cm}even.
\end{eqnarray}
The condition for perfect transmission thus becomes
\begin{eqnarray}
|\frac{t_{1}t_{3}\cdots t_{N-2}}{t_{2}t_{4}\cdots t_{N-1}}|^{2} &=&\frac{%
\Gamma ^{L}}{\Gamma ^{R}};\hspace{0.8cm}N\hspace{0.1cm}odd \\
|\frac{t_{1}t_{3}\cdots t_{N-1}}{t_{2}t_{4}\cdots t_{N-2}}|^{2} &=&\frac{%
\Gamma ^{L}\Gamma ^{R}}{4};\hspace{0.3cm}N\hspace{0.1cm}even.
\end{eqnarray}

Equation (6) states that for a chain with an odd number of sites
and mirror (inversion) symmetry ($\Gamma ^{L}=\Gamma ^{R}=\Gamma,
t_1=t_{N-1},t_2=t_{N-2},$ etc.) perfect transmission is
automatically satisfied at the middle of the
band or level group. From  the Landauer-B\"{u}ttiker formula $\cG=2e^{2}\cT %
/h$ it follows that the conductance is then quantized to the value
$2e^{2}/h. $ This is not the case when $N$ is even, however, as is
apparent from the structure of Eq. (7) and (2), yielding a
transmission coefficient $4\lambda/(1+\lambda)^2$ with
$\lambda=|\frac{2t_1t_3\cdots t_{N-1}}{\Gamma t_2t_4\cdots
t_{N-2}}|^2$ less than unity and a conductance smaller than
$2e^{2}/h$. We thus see that the even-odd feature appears in
transport in the absence of any electron correlations. Our
argument also proves that when the system is symmetric under
inversion, the state at the middle of the band is always
delocalized, regardless of the amount of disorder that respects
such symmetry condition. This is a special kind of generic
correlation in the disorder, defined by specular symmetry. Of
course equations (6) or (7) may be satisfied by a much broader set
of parameter sequences, thus defining a class of correlated
disorder constraints in which inversion symmetry is just a
particular case. For example, the simple pair-like condition
$t_1=t_2, t_3=t_4, \cdots,
t_{N-2}=t_{N-1},\Gamma^L=\Gamma^R=\Gamma$ would suffice to satisfy
the even-odd rule, being one instance within such set. Notice that
one may also find particular cases in which the rule is inverted,
such as when $t_1=t_2, t_3=t_4, \cdots, |t_{N-1}|^2=\Gamma^L
\Gamma^R/4$ with $\Gamma^L=\Gamma^R$. But they are very special
and difficult to achieve experimentally.

The above results show that the even-odd behavior found in
symmetric Hubbard chains coupled to
reservoirs,\cite{Oguri,Brouwer1} need not arise from Kondo-like or
other type of electron correlations. The results are also relevant
to the case of transport through a monatomic wire considered in
the literature.\cite{Yeyati} Then $\epsilon _{0}$ is just the site
energy of the ${\bf s}$-orbital of the noble or alkali like atoms,
while the hopping elements $t_{i}$ represent the overlap between
the nearest-neighbor ${\bf s}$ orbitals and are all equal.
Following our conclusions, if coupling to the left and right lead
is the same we then expect the conductance to be quantized, as
long as the number of atoms in the chain is odd, again exhibiting
the even-odd feature. For a wire with deformations perpendicular
to its length the even-odd character of the conductance is
preserved since the normal deformation just changes the inter-site
couplings symmetrically.

The above results are valid for any finite N, no matter how long
the chain is. Also, if the electron-electron interaction is added,
we expect them to hold as well since the on-site Coulomb
interaction introduces a self-energy term $\Sigma _{e-e}$ in the
Green's function of each site. The influence of electron-electron
interactions is then just to shift and split the resonance
position \cite{Langreth} and Eq. (2) can also be formally used in
their presence, with the formal replacement
$g_{i}^{r}=(\epsilon_{F}-\epsilon _{0}-Re\Sigma _{e-e})^{-1}$ and
the new resonance condition $\epsilon _{0}=\epsilon_{F}-Re\Sigma
_{e-e}$.

In summary, we have shown that in the absence of electron-electron
interactions a broad class of off-diagonal correlated disordered
1D samples of finite length exhibit a state with transmission
coefficient equal unity at the center of the band. The set
includes all sequences with inversion symmetry, for which perfect
transmission takes place if the number of sites is odd, while if
it is even transmission is less than one, yielding a conductance
equaling $2e^{2}/h$ in the first case, and smaller in the latter.
However, in the case of general disorder, one expects the even-odd
rule to be violated. Our results strongly suggest that previous
interpretation of this even-odd effect in terms of electron
correlations must be revised.

This work was supported in part by a Catedra Presidencial en
Ciencias and FONDECYT 1990425 (Chile), and NSF grant No.
53112-0810 of Hunan Normal University (China).


\begin{thebibliography}{15}
\bibitem {Anderson}  P. W. Anderson, Phys. Rev. {\bf 109}, 1492 (1958).
\bibitem {Mott}  N. F. Mott and W. D. Twose, Adv. Phys. {\bf 10}, 107 (1961).
\bibitem {Borland}  R. E. Borland, Proc. R. Soc. Lond. A {\bf 274}, 529
(1963).
\bibitem {Economou}  E. N. Economou and M. H. Cohen, Phys. Rev. B {\bf 4},
396 (1971).
\bibitem {Theodorou}  G. Theodorou and M. H. Cohen, Phys. Rev. B {\bf 13},
4597 (1976).
\bibitem {Hilke}  M. Hilke and J. C. Flores, Phys. Rev. B {\bf 55}, 10625
(1997).
\bibitem {Shankar}  R. Shankar and G. Murthy, Phys. Rev. B {\bf 36}, 536
(1987).
\bibitem{Mckenzie}  R. H. Mckenzie, Phys. Rev. Lett. {\bf 77}, 4804 (1996).
\bibitem {Brouwer1}  P. W. Brouwer, C. Mudry, B. D. Simons, and A. Altland,
Phys. Rev. Lett. {\bf 81}, 862 (1998).
\bibitem {Brouwer2}  P. W. Brouwer, A. Furusaki, I. A. Gruzberg, and C.
Mudry, ibid. {\bf 85}, 1064 (2000).
\bibitem {Oguri}  A. Oguri, Phys. Rev. B {\bf 59}, 12240 (1999); {\bf 63},
115305 (2001).
\bibitem {Datta}  S. Datta, {\it Electronic Transport in Mesoscopic Systems}
(Cambridge University Press, 1995), P246-273.
\bibitem {Sim}  H. -S. Sim, H. -W. Lee, and K. J. Chang, Phys. Rev. Lett.
{\bf 87}, 096803 (2001).
\bibitem {Yeyati}  A. Levy Yeyati, A. Mart\'{i}n-Rodero, and F. Flores, Phys.
Rev. B {\bf 56}, 10369\textexclamdown \textexclamdown (1997).
\bibitem {Langreth}  D. C. Langreth, Phys. Rev. {\bf 150}, 516 (1966).
\end{thebibliography}
\end{document}